\documentclass[10pt,journal,compsoc]{IEEEtran}
\usepackage{multirow}
\usepackage{stfloats}
\usepackage{algorithm}
\usepackage{algorithmic}
\usepackage{graphicx}
\usepackage{scalerel}
\usepackage{tikz}
\usetikzlibrary{svg.path}

\definecolor{orcidlogocol}{HTML}{A6CE39}
\tikzset{
    orcidlogo/.pic={
        \fill[orcidlogocol] svg{M256,128c0,70.7-57.3,128-128,128C57.3,256,0,198.7,0,128C0,57.3,57.3,0,128,0C198.7,0,256,57.3,256,128z};
        \fill[white] svg{M86.3,186.2H70.9V79.1h15.4v48.4V186.2z}
        svg{M108.9,79.1h41.6c39.6,0,57,28.3,57,53.6c0,27.5-21.5,53.6-56.8,53.6h-41.8V79.1z M124.3,172.4h24.5c34.9,0,42.9-26.5,42.9-39.7c0-21.5-13.7-39.7-43.7-39.7h-23.7V172.4z}
        svg{M88.7,56.8c0,5.5-4.5,10.1-10.1,10.1c-5.6,0-10.1-4.6-10.1-10.1c0-5.6,4.5-10.1,10.1-10.1C84.2,46.7,88.7,51.3,88.7,56.8z};
    }
}

\newcommand\orcidicon[1]{\href{https://orcid.org/#1}{\mbox{\scalerel*{
                \begin{tikzpicture}[yscale=-1,transform shape]
                \pic{orcidlogo};
                \end{tikzpicture}
            }{|}}}}

\usepackage{hyperref}

\begin{document}

\title{A Bytecode-based Approach for Smart Contract Classification}

\author{Chaochen~Shi$^{\textsuperscript{\orcidicon{0000-0002-5543-1655}}}$,
        Yong~Xiang$^{\textsuperscript{\orcidicon{0000-0003-3545-7863}}}$,~\IEEEmembership{Senior Member,~IEEE,}
        Robin~Ram~Mohan~Doss$^{\textsuperscript{\orcidicon{0000-0001-6143-6850}}}$,~\IEEEmembership{Senior Member,~IEEE,}
        Jiangshan~Yu$^{\textsuperscript{\orcidicon{0000-0001-8006-7392}}}$,~\IEEEmembership{Member,~IEEE,}
        Keshav~Sood$^{\textsuperscript{\orcidicon{0000-0001-6143-6850}}}$,~\IEEEmembership{Member,~IEEE,}
        and~Longxiang~Gao$^{\textsuperscript{\orcidicon{0000-0002-3026-7537}}}$,~\IEEEmembership{Senior Member,~IEEE}
\IEEEcompsocitemizethanks{\IEEEcompsocthanksitem C. Shi, Y. Xiang and L. Gao are with the Deakin Blockchain Innovation Lab, School of Information Technology, Deakin University, Geelong, Australia.\protect\\
E-mail: \{shicha, yxiang, longxiang.gao\}@deakin.edu.au.
\IEEEcompsocthanksitem R. Doss and K. Sood are with the Centre for Cyber Security Research and Innovation, School of Information Technology, Deakin University, Geelong, Australia.\protect\\
E-mail: \{robin.doss, keshav.sood\}@deakin.edu.au.
\IEEEcompsocthanksitem J. Yu is with the Monash Blockchain Technology Centre, Faculty of Information Technology, Monash University, Australia.\protect\\
E-mail: jiangshan.yu@monash.edu.}

\thanks{(Corresponding author: Yong Xiang.)}}

\IEEEtitleabstractindextext{%
\begin{abstract}
With the development of blockchain technologies, the number of smart contracts deployed on blockchain platforms is growing exponentially, which makes it difficult for users to find desired services by manual screening. The automatic classification of smart contracts can provide blockchain users with keyword-based contract searching and helps to manage smart contracts effectively. Current research on smart contract classification focuses on Natural Language Processing (NLP) solutions which are based on contract source code. However, more than 94\% of smart contracts are not open-source, so the application scenarios of NLP methods are very limited. Meanwhile, NLP models are vulnerable to adversarial attacks. This paper proposes a classification model based on features from contract bytecode instead of source code to solve these problems. We also use feature selection and ensemble learning to optimize the model. Our experimental studies on over 3,300 real-world Ethereum smart contracts show that our model can classify smart contracts without source code and has better performance than baseline models. Our model also has good resistance to adversarial attacks compared with NLP-based models. In addition, our analysis reveals that account features used in many smart contract classification models have little effect on classification and can be excluded.
\end{abstract}

}

\maketitle
\IEEEdisplaynontitleabstractindextext
\IEEEpeerreviewmaketitle
\IEEEraisesectionheading{\section{Introduction}\label{sec:introduction}}
\IEEEPARstart{A} smart contract is an event-driven program running on distributed ledgers. The concept of smart contract was originally introduced by Szabo~\cite{szabo1996smart}, providing a commitment defined in a digital form. As of April 2020, the number of smart contracts on Ethereum~~\cite{buterin2014next} exceeds two million~\cite{etherscan}. As the number of contracts increases, how to help users find the services they need in massive contracts has become an important issue. The primary query APIs of smart contracts provided by blockchain platforms are based on contract address, block number, transaction hash, and timestamp. Some commercial tools such as Google Bigquery~\cite{bigquery} and Dfuse~\cite{dfuse} provide SQL and GraphQL supported blockchain databases to realize complex queries on row data. However, as blockchain platforms are gradually evolving into distributed data centers, users desire a more convenient searching experience, e.g., searching by keywords~\cite{jiang2020searchain} or categories~\cite{huang2017towards}. An essential step to conduct such searching is labeling smart contracts accurately. Currently, the identification of smart contracts relies on manual labeling, which is costly and inefficient. Therefore, it is necessary to design an effective classification model to classify and label existing or newly uploaded contracts automatically. The goal of the classification is as follows:

The dataset is defined as $\{D_i,y_j\}$, where $D_i$ refers to a smart contract and $y_j$ belongs to $Y$ which is a predefined collection of $k$ categories, $Y=\{y_1, y_2, \ldots, y_k\}$. The goal is to learn a mapping function $h$ which maps an input $D_i$ to the category $y_j$ it belongs to.

\begin{figure*}[!t]
  \centering
  \includegraphics[width=\linewidth]{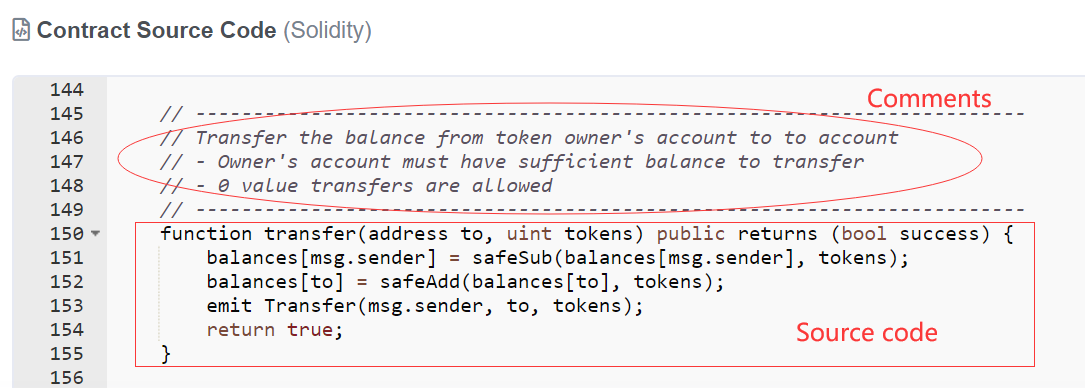}
  \caption{The composition of the smart contract.}
  \label{fig:code}
\end{figure*}

As shown in Fig.~\ref{fig:code}, a smart contract consists of source code and comments. Currently, the prevalent method to classify smart contracts is using NLP techniques such as the attention-based LSTM network~\cite{tang2015effective} to capture the semantic features from source code and comments (described in Section 2). However, there are two main problems in the existing NLP-based models:
\begin{enumerate}[]
\item NLP-based models have limited application scenarios. They can only classify open-source contracts~\cite{tian2020smart}. However, open-source code is not mandatory for contract developers. Less than 4\% contracts are open-source on Ethereum~\cite{etherscan}, which means that NLP models cannot classify more than 96\% of smart contracts;
\item NLP-based models can be easily attacked by adversarial examples~\cite{zhang2020adversarial}. Context features like comments, variable names, and function names can be easily modified in the source code without changing the logic of the code. Developers can write their source code in different ways to perform the same functions; however, any additions, deletions, or modification of the source code may fool the NLP-based classifier. 
\end{enumerate}

To solve these problems, we need models which can classify smart contracts without the source code. Inspired by the wide use of bytecode features in other areas of smart contracts, such as contract vulnerability and fraud detection, we found that bytecodes can reflect the functional features from the logical aspect. We expect that bytecode features can also be successfully applied to smart contract classifications.

In this paper, we propose a multi-classification model for smart contracts based on bytecode features. Compared with NLP-based models, our bytecode-based approach can effectively classify smart contracts without source code, which significantly expands the application scenarios. In addition, adversarial attacks against the source code have little effect on the bytecode-based model because the attacked semantic information is discarded after being compiled. Considering that the categories of smart contracts are unevenly distributed in the blockchain platforms, we propose employing feature selection (Binary Particle Swarm Optimization~\cite{kennedy1997discrete}) and ensemble learning techniques (Adaboost~\cite{sun2007cost}) to solve the problem of data imbalance in our models. This paper focuses on the Ethereum platform, but the approach can easily be expanded to other blockchain platforms. The key contributions of this paper are as follows.
\begin{enumerate}[]
\item We propose a bytecode-based approach, which is the first approach to classify smart contracts when their source code is intentionally hidden.
\item We demonstrate that our bytecode-based approach has better resistance to adversarial attacks than state-of-the-art source code-based approaches.
\item We prove that feature selection and ensemble learning are competitive alternatives to solve data imbalance problems in smart contracts classification.
\item We determine that account features have little effect on classifications compared with code features and explain the reason. It has guiding significance for future research on the classification model of smart contracts.
\end{enumerate}

\section{Background and Related Work}

\subsection{Background}
Ethereum is one of the most popular programmable blockchain with a built-in Turing-complete instruction set. Users can develop customized cryptocurrencies or decentralized applications (Dapps) built on smart contracts on the Ethereum platform. As the core of Ethereum, the Ethereum virtual machine (EVM) can compile high-level programming languages such as Solidity into bytecode. The bytecode consists of a series of bytes, and each byte refers to a specific operation represented by a corresponding mnemonic form predefined in the Ethereum yellow paper~\cite{wood2014ethereum}. For example, the mnemonic of value 0x01 is ADD, which means the add operation. These mnemonic forms are called opcodes, which reflect the operational logic of programs directly from EVM level. Table~\ref{tab1} lists some frequently-used opcodes and their meanings.
\begin{table}
\caption{Examples of Ethereum opcodes.}\label{tab1}
\centering
\vspace{3pt} \noindent
\begin{tabular}{p{50pt}p{50pt}p{120pt}}
\hline
\parbox{39pt}{\centering
{Value}
} & \parbox{48pt}{\centering
{Mnemonic}
} & \parbox{35pt}{\centering
{Description}
} \\
\hline
\parbox{50pt}{\centering
{0x06}
} & \parbox{48pt}{\centering
{MOD}
} & \parbox{120pt}{
{Modulo addition operation.
}
} \\
\parbox{50pt}{\centering
{0x0a}
} & \parbox{48pt}{\centering
{EXP}
} & \parbox{120pt}{
{Exponential operation.
}
}  \\
\parbox{50pt}{\centering
{0x10}
} & \parbox{48pt}{\centering
{LT}
} & \parbox{120pt}{
{Less than operation}
} \\
\parbox{50pt}{\centering
{0x33}
} & \parbox{48pt}{\centering
{CALLER}
} & \parbox{120pt}{
{Get the caller address.}
} \\
\parbox{50pt}{\centering
{0x5a}
} & \parbox{48pt}{\centering
{GAS}
} & \parbox{150pt}{
{Get the amount of available gas.}
} \\
\parbox{50pt}{\centering
{0x60}
} & \parbox{48pt}{\centering
{PUSH1}
} & \parbox{150pt}{
{Place a 1 byte item on the stack.}
} \\
\parbox{50pt}{\centering
{0x54}
} & \parbox{48pt}{\centering
{SLOAD}
} & \parbox{150pt}{
{Load the first word from storage.}
} \\
\hline
\end{tabular}
\end{table}

Based on the EVM, developers can deploy smart contracts on the Ethereum platform easily. The process can be divided into three steps: first, use a high-level language like Solidity to write the smart contract source code; second, compile the source code into bytecode through EVM; and finally, deploy the compiled contract through Ethereum clients.

Every user of Ethereum can hold an account. An Ethereum account has a 20-byte address, including four unique fields: nonce, balance, contract bytecode (if any), and storage (usually empty). Only contract accounts have code fields which store codeHash (the hash value of the EVM code for this account). This field cannot be modified after creation, which means that the smart contract is immutable. When the contract account receives a message, the contract is activated. This allows it to read and write to the internal storage, send messages out, or create a new contract. We use both bytecode features and account information in training the smart contract classifier.

\subsection{Related Work}
There are few studies on the classification of smart contracts. Huang et al.~\cite{huang2017towards} have introduced a smart contract classification method based on the word embedding model. This method captures the semantics of the contract source code through the LSTM network and obtains word vectors. Finally, word vectors and account characteristics are input into the feedforward neural network; the probability distribution of the category labels is output. Gang et al.~\cite{tian2020smart} have proposed a novel classification model called SCC-BiLSTM. It employs the Gaussian LDA (GLDA) model and attention mechanism to improve the classifier's performance. This model solves the sparse semantic problem of annotations in the source code, and the attention mechanism is used to capture vital code features. The experimental results show that this model achieves superior effectiveness on smart contract classification tasks, but it still relies on open-source contracts.

Studies have used bytecode or opcode to analyze smart contracts. Oyente~\cite{luu2016making} is a symbolic execution tool released by Melonport to detect potential security vulnerabilities such as reentrancy, timestamp dependence, and logic errors in smart contracts. Oyente works directly with the EVM bytecode and opcodes without access to high-level programming languages like Solidity or Python. The research by Chen et al.~\cite{chen2018detecting} has used features extracted from bytecode to detect Ponzi schemes in Ethereum smart contracts. This model extracts features from bytecode in manually labeled contract samples and trains the regression tree model with the XGBoost~\cite{chen2016xgboost} algorithm. The most significant innovation is that by using this bytecode-based model Ponzi schemes can be detected once contracts are created. Chen et al.~\cite{chen2018understanding} conducted an investigation on Ethereum through graph analysis. They collected all transaction data by customizing Ethereum client using opcodes. Barati et al.~\cite{bar20} show that some data privacy rules can be translated into smart contracts and appear as opcodes to verify the way providers operate user data automatically. 

Unlike source code-based approaches, we use features from contract bytecode to train the classification model. Since bytecode is immutable and is open to access, the bytecode-based classifier is universal to all contracts no matter they are open-source or not. This is the main difference between our approach and NLP-based approaches.

\section{Proposed Methodology}
\subsection{Framework}
\begin{figure*}[tb]
  \centering
  \includegraphics[width=\linewidth]{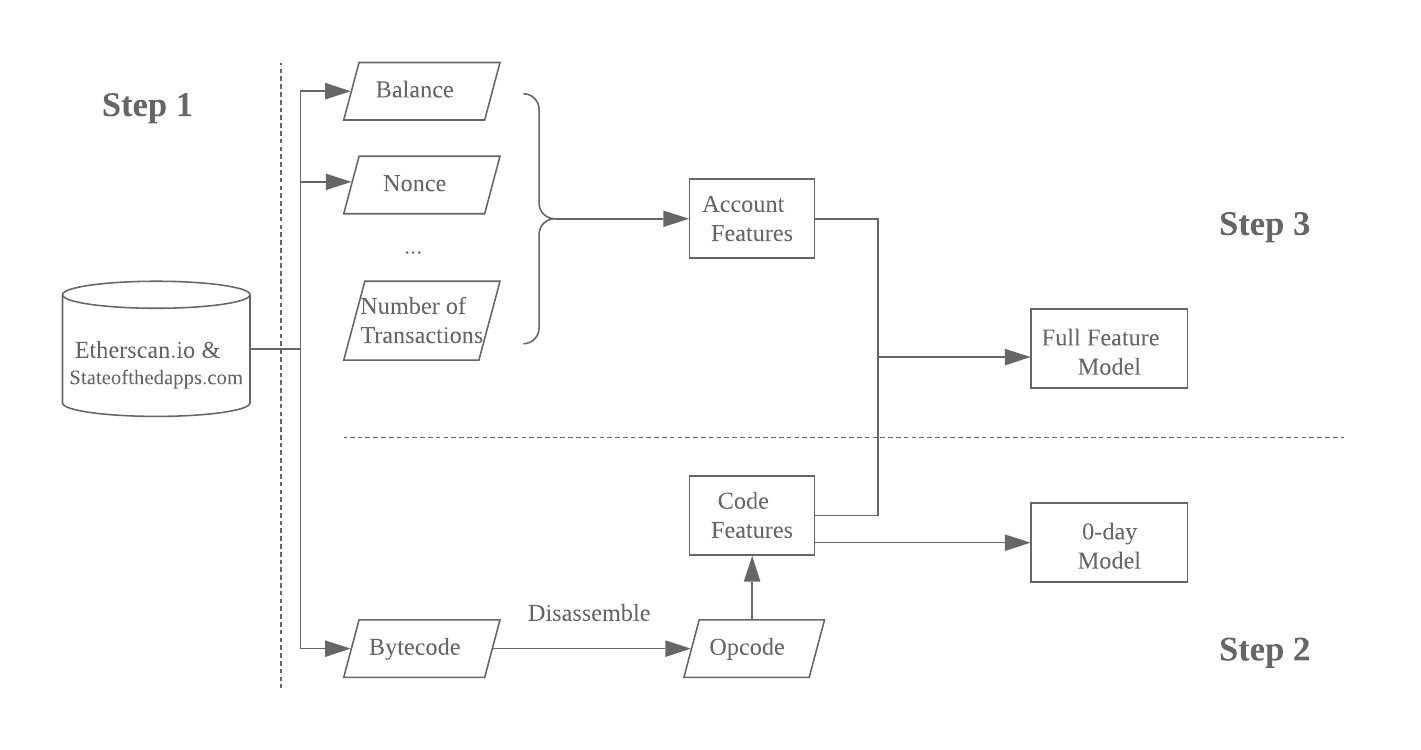}
  \caption{The framework of the smart contracts classification approach.}
  \label{fig:framework}
\end{figure*}

The overall framework is illustrated as Fig~\ref{fig:framework}. We first collected verified smart contracts by crawling from Ethereum explore (\textit{etherscan.io} and \textit{stateofthedapps.com}), including the contract bytecode and related account information. The second step converts the bytecode into opcodes and extracts the code features to train the 0-day model, which classifies contracts once they are uploaded. The third step integrates the contract behavior features from the transaction history to train the full-feature model. To solve the problems of feature redundancy and sample imbalance in the model training, we also propose an ensemble learning-based multi-classification algorithm with a binary particle swarm optimization (BPSO) method. 

\subsection{Data}
\begin{figure}[tb]
  \centering
  \includegraphics[width=\linewidth]{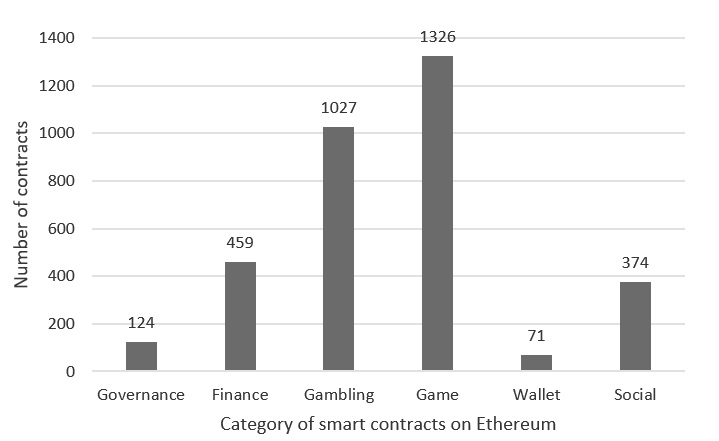}
  \caption{The distribution of collected smart contracts.}
  \label{fig:categories}
\end{figure}

To train and test our model, we collected 11,000 smart contracts of top 100 Ethereum Dapps ranked by their user activities (unique source addresses in transactions to DApp contracts) over the past 30 days, as of May 1, 2020. All smart contracts are collected from Ethereum explores \textit{etherscan.io} and \textit{stateofthedapps.com} through web crawlers. After deleting duplicate contracts and contracts which have never been triggered, 3,381 contracts are left, and 1,501 contracts of them are open-source. Each contract contains full information, including the bytecode and account information. We also collected all of the transaction histories of these contracts, such as the number of transactions and the amount of transferred Ether for further feature extraction. The collected contracts are manually divided into six categories: \textit{Governance, Finance, Gambling, Game, Wallet, Social} according to the Dapps to which they belong. The distribution of the collected contracts is shown in Fig~\ref{fig:categories}. The imbalance ratio of the samples is 19, which is similar to the current Ethereum environment; game and gambling contracts appear more frequently than other categories.

\subsection{Feature Extraction}
Feature extraction and selection are key upstream capabilities for building a high-performance classifier. Previous work, e.g., ~\cite{luu2016making,pham2016anomaly} has extracted code features mainly from bytecode for vulnerability detection and pattern recognition. In other proposals, e.g., ~\cite{tian2020smart,8525395}, account and transaction information have been selected as account features to characterize contracts. In our work, we build a 0-day model that is based on code features to classify contracts as early as day 0. This is possible because the code features are available immediately and are immutable once the contracts are uploaded. We integrate code features and account features to train the full-feature model to improve the classification accuracy for the already deployed contracts.

\subsubsection{Code Features}
As the main body of a smart contract, the bytecode is stored as a string of hexadecimal numbers with the contract account in a Merkle Patricia tree. Unlike source code, bytecode is transparent and can be easily obtained from every contract. As mentioned in Section 1, each byte represents a certain opcode, so we disassemble the bytecode into equivalent opcode with \emph{evmdis\footnote{https://github.com/Arachnid/evmdis, accessd May 1, 2020}} to facilitate the feature extraction. The opcode features can be directly used in classification without any modifications because they reflect all of the logical behaviors of contracts~\cite{atzei2017survey} from the perspective of the EVM.

After disassembling bytecodes into opcodes, the frequency of each kind of opcode is calculated and regarded as a feature. Please note that for some opcodes with the same functions, we merge them into one category. For example, both DUP1 and DUP2 are considered as DUP; both PUSH1 and PUSH2 are considered as PUSH. Finally, we find 61 different kinds of opcodes from all 1,501 contracts, which means the dimension of the code feature is 62, including the size of the bytecode. Table~\ref{tab2} shows the top 10 code features (except size feature) ranked by their average values in three categories.

\begin{table*}[htb]
\caption{Top 10 code features ranked by average values.}\label{tab2}
\centering
\begin{tabular}{c|cc|cc|cc}
\hline
\multirow{2}{*}{Rank} & \multicolumn{2}{c|}{Game contracts} & \multicolumn{2}{c|}{Social contracts} & \multicolumn{2}{c}{Financial contracts} \\ \cline{2-7} 
                      & Feature         & Value (avg)        & Feature          & Value (avg)         & Feature           & Value (avg)          \\ \hline
1                     & PUSH            & 134.18            & PUSH             & 84.13              & PUSH              & 57.36               \\
2                     & DUP             & 97.44             & JUMP             & 52.12              & JUMP              & 50.21               \\
3                     & SWAP            & 89.15             & DUP              & 45.32              & SWAP              & 18.35               \\
4                     & JUMP            & 53.24             & MSTORE           & 32.87              & RETURN            & 14.23               \\
5                     & POP             & 39.23             & SSTORE           & 31.25              & DUP               & 9.09                \\
6                     & RETURN          & 11.21             & SWAP             & 7.62               & MUL               & 4.23                \\
7                     & MLOAD           & 4.35              & CALL             & 3.38               & MSTORE            & 2.21                \\
8                     & CALL            & 2.31              & POP              & 1.09               & SUB               & 0.77                \\
9                     & MSTORE          & 0.89              & AND              & 0.43               & STOP              & 0.34                \\
10                    & ADD             & 0.72              & ISZERO           & 0.31               & CREATE            & 0.15                \\ \hline
\end{tabular}
\end{table*}

According to Table~\ref{tab2}, the distributions of code features are different among the three categories. The most frequently used features are PUSH, DUP, SWAP, and JUMP. These features all relate to stack operations. That is because almost any operation, such as defining variables and functions, involves stack operations on the EVM. We also found that MSTORE and MLOAD are more frequently used in game contracts than in others. This outcome is reasonable because some game data needs to store and load from memory. Other categories of contracts also have characteristic feature distributions that reflect their unique characteristics. Thus, we believe code features can be used in contract classification. Although there are some common frequently appearing opcodes among different categories, we still use all 61 opcodes as features because they may have hidden connections with each other and cannot be excluded by a simple standard.

\subsubsection{Account Features}
Account features are selected from account attributes and related transaction history information. These features are only available after the deployment of contracts and may change over time, reflecting how contracts work in a real environment. Thus, we can extract these features from already deployed contracts and combine them with code features to train a full-feature model that classifies deployed contracts.

Previous research~\cite{tian2020smart,pham2016anomaly,8525395} provides a variety of account features. From these, we select features to model smart contracts as follows:
\begin{itemize}
\item \textbf{Balance:} the balance of contract account, measured by \textbf{wei}. 
\item \textbf{Nonce:} the nonce records the sequence of contract creation.
\item \textbf{Nbr\_trans\_act} and \textbf{Nbr\_trans\_psv:} the number of active and passive transactions involving the contract.
\item \textbf{Eth\_in} and \textbf{Eth\_out:} the total amount of income and output Ether of the contract.
\item \textbf{Eth\_avg} and \textbf{Eth\_sdev:} the mean and standard deviation of the Ether transferred by the contract.
\item \textbf{Lifetime:} the time gap between the initial and the last transaction.
\item \textbf{Trs\_gap\_avg} and \textbf{Trs\_gap\_sdev:} the average and standard deviation of the time gap between every two transactions.
\item \textbf{Nbr\_addr:} the total number of addresses the contract interacted with.
\end{itemize}

\subsection{Feature Selection}
According to the statistics from \textit{stateofthedapps.com}, the distribution of the smart contract categories on Ethereum is very uneven. The number of game and gambling contracts is much larger than other contracts, while wallet and governance contracts are rare. Thus, the smart contract classification problem can be regarded as a multi-classification problem on an imbalanced data set. Traditional classification algorithms such as Decision Tree, K-Nearest Neighbor, and Support Vector Machine present challenges in achieving the desired performance on an imbalanced data set because of their bias towards the majority class. It may treat the minority class samples as noise~\cite{sun2007cost}, which results in the poor classification performance of the minority class.

To improve the classification performance, We integrate feature selection in our classification model. The feature selection process can eliminate the irrelevant and redundant features to reduce the noise in the sample space~\cite{chandrashekar2014survey}, thereby improving the classification performance of minority classes. In addition, feature selection also helps us find critical features and hidden relationships among a massive number of original features and decreases the time complexity.

When we eliminate the irrelevant features, there is also a risk to the potential loss of useful information because the feature selection procedure may alter the original data distribution~\cite{sun2015novel}. We prefer employing warpper methods rather than filter methods such as Mutual Information~\cite{guyon2003introduction} or Relief-based algorithms~\cite{urbanowicz2018relief} to imbalanced data classification, since the correlation between features and targets is not clear. BPSO~\cite{kennedy1997discrete} is a stochastic evolutionary algorithm which is widely used for solving optimization problems in binary space. Compared with other wrapper methods such as Genetic algorithm, Differential Evolution, etc., the complexity of BPSO is much lower since it does not contain crossover and mutation operations. In this paper, we choose BPSO as the feature selection method. Our method follows the original BPSO algorithm and only changes the particle representation and fitness values. We encode binary particles as a multi-dimensional vector with values $[1,0]$ and each bit of the vector represents a feature which is selected (value 1) or not (value 0). The fitness value of a particle is usually the classification accuracy of the sample subset indicated by the particle. Here, we choose the normalized AUC\_area shown as Eq.~(\ref{eq}) (decribed in Section 4.1) as the fitness value instead.

\subsection{Classification Model}
Ensemble learning~\cite{dietterich2002ensemble} is a machine learning method that uses a specific rule to combine multiple classifiers as a collection to achieve better predictive performance than an individual classifier. The idea of ensemble learning is that even if a weak classifier obtains an incorrect prediction, other classifiers can correct it. Adaboost.M1~\cite{freund1996experiments} is a typical ensemble learning algorithm that has been widely used to solve multi-classification problems because of its good performance, low complexity, and good resistance to overfitting. Adaboost.
M1 creates a simple weak learner for each feature. Weak learners do not need high accuracy in the initial stages, as long as their accuracy is higher than random classification. The weight of the correctly classified samples decreases, and the weight of the incorrectly classified samples remains unchanged after each iteration. Suppose $m$ is the number of samples and $Y$ is the collection of $k$ categories, $Y = \{c_1, c_2, \ldots, c_k\}$. Then the classification error rate is
\begin{equation}
\label{eq1}
\varepsilon = \sum_{i=1}^mD(i)[y_i \ne h(x_i)],
\end{equation}
where $D(i)$ is the weight of sample $x_i$, $y_i\in Y$, and $h$ is the weak hypothesis $h: x_i \rightarrow Y$. Setting a parameter
\begin{equation}
\label{eq2}
\beta = \frac{\varepsilon}{1-\varepsilon},
\end{equation}
the weight would be updated as
\begin{equation}
\label{eq3}
D_{t+1}(i)=D_{t}(i)\beta_t^{1-[y_i \ne h_t(x_i)]},
\end{equation}
where $t$ is the current number of iterations. In this way, the distribution of samples becomes more balanced after each iteration. Finally, we can obtain a strong classifier with a superior predictive performance by combining the weak learners obtained in iterations. The final strong hypothesis $H(x)$ is
\begin{equation}
\label{eq4}
H(x)=\mathop{\arg\max}\limits_{y\in Y}(\sum_{t=1}^T\ln(\frac{1}{\beta_t})[y = h_t(x)]),
\end{equation}
where $T$ is the total number of iterations.

Adaboost.M1 requires relatively high-performance weak learners. We choose C4.5~\cite{quinlan2014c4} as the algorithm of the weak learner based on two reasons: 
\begin{enumerate}[]
\item There are numerous missing values in our code features. Thus, C4.5 is suitable for our case as it has good performance and low sensitivity to missing values.
\item Although all of the binary classification algorithms can be expanded to multi-classification versions via the OvO or OvA strategy~\cite{fernandez2013analysing}, this significantly increases the complexity of the algorithm. So, we choose C4.5, which can be directly used in multi-classification problems.
\end{enumerate}

We use the BPSO algorithm for feature extraction and then put the sample subset selected from $m$ samples by the particles into the Adaboost.M1 algorithm. To compute the AUC\_area as Eq.~(\ref{eq}), the classifier needs to output a $k$-dimensional probability vector for each sample in $k$-class classification. Values in the probability vector are the probabilities of a sample belongs to each class. So there are two probability matrixes $Weak\_score$ and $Strong\_score$ with size $m\times k$ belonging to the weak learner and the final strong classifier respectively. Since $\varepsilon_t$ and $Weak\_score_t$ are updated in each iteration, the $Strong\_score$ could be the average of weighted $Weak\_score$:
\begin{equation}
\label{eq5}
Strong\_score =\sum_{t=1}^T(\frac{1-\varepsilon_t}{\sum_{t=1}^T1-\varepsilon_t}Weak\_score_t).
\end{equation}

Algorithm~\ref{alg:Framwork} presents the pseudocode of the whole classification algorithm, including the training and classification process. In additon to $H(x)$, we can also obtain the prediction results and the optimal subset of features $S_{best}$. The notations used in the algorithm are listed in Table~\ref{notation}.
\begin{table}[t]
\centering
\caption{The notations used in algorithm 1.}
\label{notation}
\begin{tabular}{l|l}
\hline
Notation     & Explanation                                                 \\ \hline
$Data$       & the training set;                                           \\
$m$          & the size of $Data$;                                         \\
$Y$          & the collection of $k$ categories, $Y=\{c_1, c_2, \ldots, c_k\}$; \\
$T$          & the number of iterations of Adaboost.M1;                    \\
$N$          & the scale of particles;                                     \\
${G\_max}$   & the generation limit of BPSO;                               \\
$w$          & the inertia weight of BPSO;                                 \\
$c_1$, $c_2$ & the acceleration factors of BPSO.                           \\ 
$H(x)$       & the strong hypothesis obtained by ensemble learning;        \\
$S_{best}$   &  the optimal subset of features.        \\\hline
\end{tabular}
\end{table}

 \begin{algorithm}[t]
 \caption{Framework of the BPSO-Adaboost algorithm.}
 \label{alg:Framwork}
 \begin{algorithmic}[1]
 \renewcommand{\algorithmicrequire}{\textbf{INPUT:}}
 \renewcommand{\algorithmicensure}{\textbf{OUTPUT:}}
 \REQUIRE
      $Data$ with $m$ samples and $k$ categories; $T$, $N$, ${G\_max}$ and $c_1$, $c_2$.
 \ENSURE
      $H(x)$, $S_{best}$ and the classification results;
  \STATE Initialize each particle randomly as~\cite{kennedy1997discrete}, $G=1$;
    \WHILE{(Number of generations $G < G\_max \parallel$ BPSO converged)}
    \FOR{$i=1$ to $N$} 
    \STATE Select sample subset $Data_i$ according to particle $i$;
    \STATE Initialize distribution $D_1(i) \leftarrow {1}/{m}$;
    \FOR{$t=1$ to $T$}
    \STATE Train C4.5 classifier with $D_t(i)$ to obtain weak hypothesis $h_t:Data_i \rightarrow Y$ and $Weak\_score_t$;
    \STATE Compute the classification error rate $\varepsilon_t$ as Eq.~(\ref{eq1});
    \STATE Set parameter $\beta_t$ as Eq.~(\ref{eq2});
    \STATE Update $D_{t+1}(i)$ as Eq.~(\ref{eq3});
    \ENDFOR 
    \STATE Obtain $H(x)$ as Eq.~(\ref{eq4});
    \STATE Compute $Strong\_score$ as Eq.~(\ref{eq5});
    \STATE Set $fitness(x_i) \leftarrow AUC\_area_i$, update the fitness value $p_{best}$ of each particle and the global fitness value $g_{best}$ based on $fitness(x_i)$;
    \STATE Update the position and velocity of particles with parameter $w$, $c_1$ and $c_2$;
    \ENDFOR
    \STATE $G \leftarrow G+1$;
    \ENDWHILE
  \end{algorithmic}
  \end{algorithm}

\section{Experiments and Analysis}
\subsection{Evaluation Metrics}
For imbalanced data, the Receiver Operating Characteristic (ROC) curve~\cite{fawcett2006introduction} is a well-recognized evaluation metric of classifier performance. ROC curve comes from confusion matrix as Table~\ref{tab3}, taking FPR~(${FP}/(TN+FP)$) as X-axis and TPR~(${TP}/(TP+FN)$) as Y-axis. However, the ROC curve can not quantitatively evaluate the performance of classifiers. Thus the area under the ROC curve (AUC) is widely used as the evaluation metric. The bigger the AUC is, the better the classifier performance is.

Traditional AUC values can only be used in binary classifiers. For n-class classification problems, we can combine classes in pairs and find the AUC of each pair individually. Finally, there are $C_n^2$ AUC values. We put all AUC values in a polar coordinate system and calculate the area of the graph covered by all AUC values as the metric, called AUC\_area~\cite{hassan2010novel}. The larger AUC\_area means better classification performance. Assuming there are $q$ AUC values $r_1, r_2, \ldots, r_q$ respectively where $q=C_n^2$, the normalized AUC\_area is:
\begin{equation}
\label{eq}
AUC\_area=\frac{1}{q}(\sum_{i=1}^{q-1}(r_i\times{r_{i+1}})+r_q\times{r_1})
\end{equation}

AUC\_area is sensitive to categories which have poor AUC values. If there is a poor AUC, the AUC\_area will also be poor. Therefore, a classification model needs to obtain high AUC values on all categories to keep a high AUC\_area. Compared with average AUC value, AUC\_area is more suitable for our case since it has no bias toward the majority category. In this paper, we use normalized AUC\_area, accuracy, and Micro-F1 score as evaluation metrics.

\begin{table}[]
\centering
\caption{Confusion matrix of binary classification.}
\label{tab3}
\begin{tabular}{cccc}
                                                                              &          & \multicolumn{2}{c}{Actual Values} \\ \cline{2-4} 
                                                                              &          & Positive        & Negative        \\ \cline{2-4} 
\multirow{2}{*}{\begin{tabular}[c]{@{}c@{}}Predictive \\ Values\end{tabular}} & Positive & TP              & FP              \\
                                                                              & Negative & FN              & TN              \\ \cline{2-4} 
\end{tabular}
\end{table}

\subsection{Experiment Settings}
We train and test our 0-day model and full-feature model (mentioned in section 3.2) with 10-fold cross-validation~\cite{kohavi1995study} on all 3,381 contracts. We specify $T = 30$ and specify BPSO related parameters as standard PSO algorithm settings, given in Table~\ref{tabpara}. To evaluate the effect of feature selection and ensemble learning in smart contract classification, we compared our algorithm with C4.5, Adaboost.M1 and BPSO-based C4.5. 
\begin{table}[]
\centering
\label{tabpara}
\caption{BPSO parameters used in our model.}
\begin{tabular}{cc}
\hline
Parameter & Value \\ \hline
$N$       & 30    \\
$G_{max}$ & 50    \\
$w$       & 0.73  \\
$c_1$     & 1.5   \\
$c_2$     & 1.5   \\ \hline
\end{tabular}
\end{table}

We train models with the same data set and test them on different occasions to compare the performance and robustness between our approach and state-of-the-art NLP-based approaches. Please note that even the data sets they use are the same. The code features they use are from bytecode and source code, respectively. Thus we train models on 1,301 verified contracts that have both source code and bytecode. We test these models on three different test sets: verified contracts (by employing 10-fold cross-validation), 1,880 unverified contracts, and 1,020 contracts with adversarial source code. Adversarial source code means each comment, variable name and function name of the source code is attacked by one of the four operations randomly: add, drop, swap, and replacement. It is similar to real-world attack settings. 
\subsection{Performance Evaluation}
\subsubsection{Performance comparison between 0-day model and full-feature model}
Table~\ref{tab4} and~\ref{tab5} show the performance of each algorithm on the 0-day model and full-feature model respectively over our data set. To compare the classification performance of different algorithms intuitively, Fig.~\ref{fig:0-day} and~\ref{fig:full} show the polar graphs of the AUC values of these algorithms. In the following figures, six categories: \textit{Governance, Finance, Gambling, Game, Wallet, Social} are numbered from 1 to 6 in order. The results show that the gap between the two models are tiny. In fact, 23 code features and only 2 account features, (\textbf{Balance} and \textbf{Trs\_gap\_avg}) were selected in the best feature subset $S_{best}$ of full-feature model. It means account features have little impact on classification performance. That is reasonable because account features are not stable since they change over time and can be influenced by many external factors. For example, market volatility and policy changes may lead to a substantial increase or decrease in \textbf{Eth\_avg} and \textbf{Eth\_sdev}. Thus account features are not robust enough to be as key features. In most cases, our 0-day model is sufficient for both newly uploaded and deployed contracts because contract bytecode is immutable after uploading. 

\begin{table*}[]
\centering
\caption{Performance of different algorithms on 0-day model.}
\label{tab4}
\resizebox{\textwidth}{!}{
\begin{tabular}{cccccccccc}
\hline
\multirow{2}{*}{Algorithm} & \multirow{2}{*}{AUC\_area} & \multirow{2}{*}{Micro-F1} & \multicolumn{1}{c}{\multirow{2}{*}{\begin{tabular}[c]{@{}c@{}}Overall\\ Accuracy\end{tabular}}} & \multicolumn{6}{c}{Accuracy for each category}           \\ \cline{5-10} 
                           &                          & \multicolumn{2}{c}{}                                                    & governance & Finance & Gambling & Game  & Wallet & Social \\ \hline
BPSO-Adaboost              & 0.923                    & 0.955   & 0.932                                                                       & 0.917     & 0.923   & 0.910    & 0.951 & 0.885  & 0.919  \\
Adaboost                   & 0.894                    & 0.878   & 0.904                                                                       & 0.792     & 0.897   & 0.912    & 0.946 & 0.779  & 0.871  \\
BPSO-C4.5                  & 0.829                    & 0.911   & 0.851                                                                       & 0.788     & 0.834   & 0.864    & 0.887 & 0.795  & 0.828  \\
C4.5                       & 0.797                    & 0.721   & 0.780                                                                       & 0.745     & 0.803   & 0.852    & 0.893 & 0.731  & 0.778  \\
\hline
\end{tabular}}
\end{table*}

\begin{table*}[htb]
\centering
\caption{Performance of different algorithms on a full-feature model.}
\label{tab5}
\resizebox{\textwidth}{!}{
\begin{tabular}{cccccccccc}
\hline
\multirow{2}{*}{Algorithm} & \multirow{2}{*}{AUC\_area} & \multirow{2}{*}{Micro-F1} & \multicolumn{1}{c}{\multirow{2}{*}{\begin{tabular}[c]{@{}c@{}}Overall\\ Accuracy\end{tabular}}} & \multicolumn{6}{c}{Accuracy for each category}           \\ \cline{5-10} 
                           &                          & \multicolumn{2}{c}{}                                                                            & governance & Finance & Gambling & Game  & Wallet & Social \\ \hline
BPSO-Adaboost              & 0.931                    & 0.964                    & 0.939                                                                       & 0.922     & 0.928   & 0.915    & 0.960 & 0.891  & 0.924  \\
Adaboost.M1                   & 0.904                    & 0.893                    & 0.918                                                                       & 0.799     & 0.904   & 0.911    & 0.934 & 0.792  & 0.841  \\
BPSO-C4.5                  & 0.845                    & 0.920                     & 0.877                                                                       & 0.796     & 0.857   & 0.875    & 0.899 & 0.795  & 0.833  \\
C4.5                       & 0.803                    & 0.738                    & 0.774                                                                     & 0.747     & 0.812   & 0.861    & 0.897 & 0.723  & 0.769  \\
\hline
\end{tabular}}
\end{table*}

\begin{table*}[htb]
\caption{The performance of the BPSO-Adaboost and SCC-BiLSTM algorithms on different test sets.}
\label{tab6}
\resizebox{\textwidth}{!}{
\begin{tabular}{cccc|ccc|ccc}
\hline
\multirow{2}{*}{Algorithm} & \multicolumn{3}{c|}{Verified Contracts} & \multicolumn{3}{c|}{Unverified Contracts} & \multicolumn{3}{c}{Adversarial Examples} \\ \cline{2-10} 
                           & AUC\_area     & Micro-F1    & Accuracy    & AUC\_area     & Micro-F1     & Accuracy     & AUC\_area    & Micro-F1    & Accuracy    \\ \hline
BPSO-Adaboost              & 0.916       & 0.943       & 0.920       & 0.904       & 0.938        & 0.934        & 0.912      & 0.946       & 0.929       \\
SCC-BiLSTM                 & 0.925       & 0.957       & 0.933       & 0.242       & 0.457        & 0.489        & 0.389      & 0.638       & 0.664       \\ \hline
\end{tabular}}
\end{table*}

\begin{figure}[tb]
  \centering
  \includegraphics[width=\linewidth]{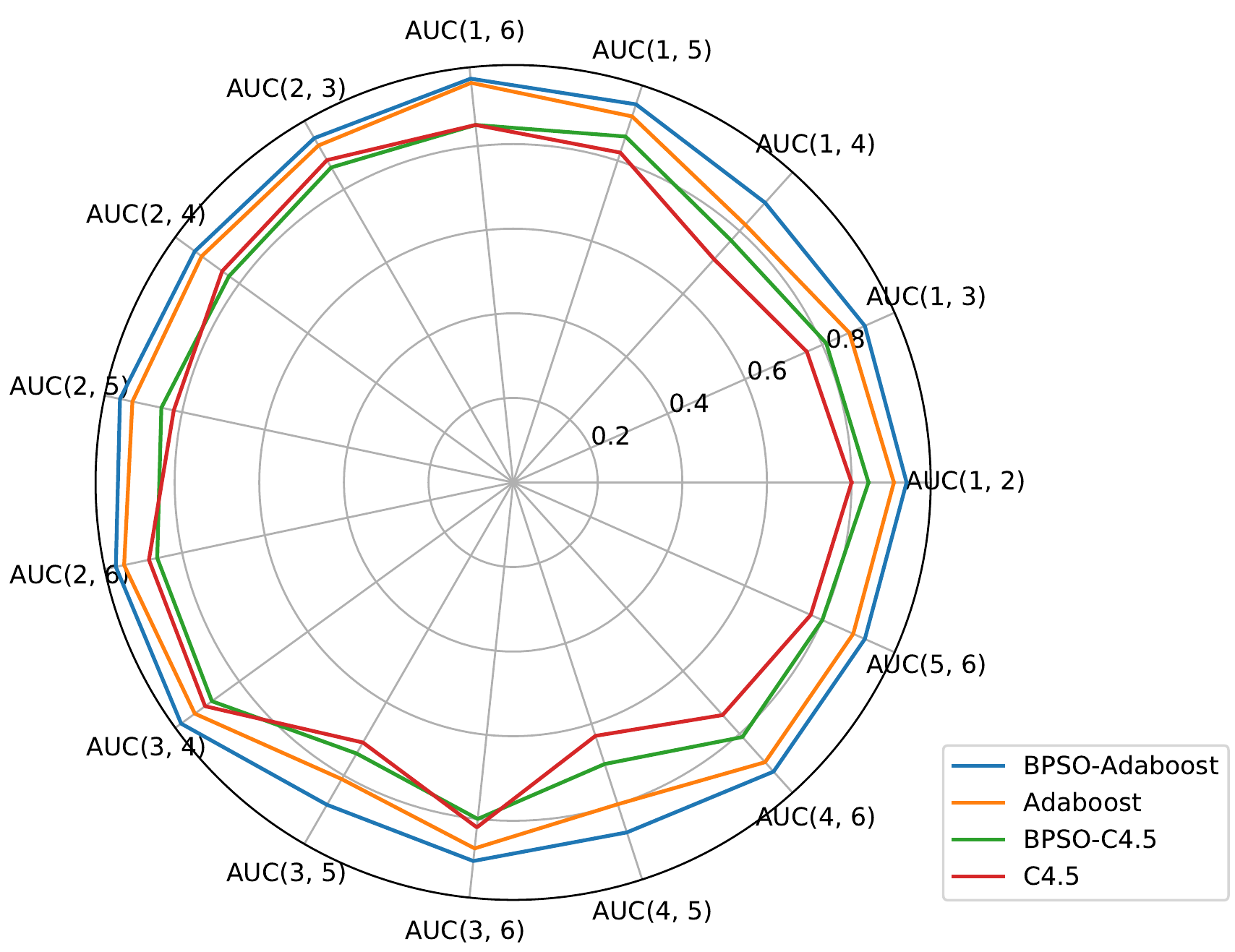}
  \caption{The AUC\_area of algorithms on 0-day model}
  \label{fig:0-day}
  \includegraphics[width=\linewidth]{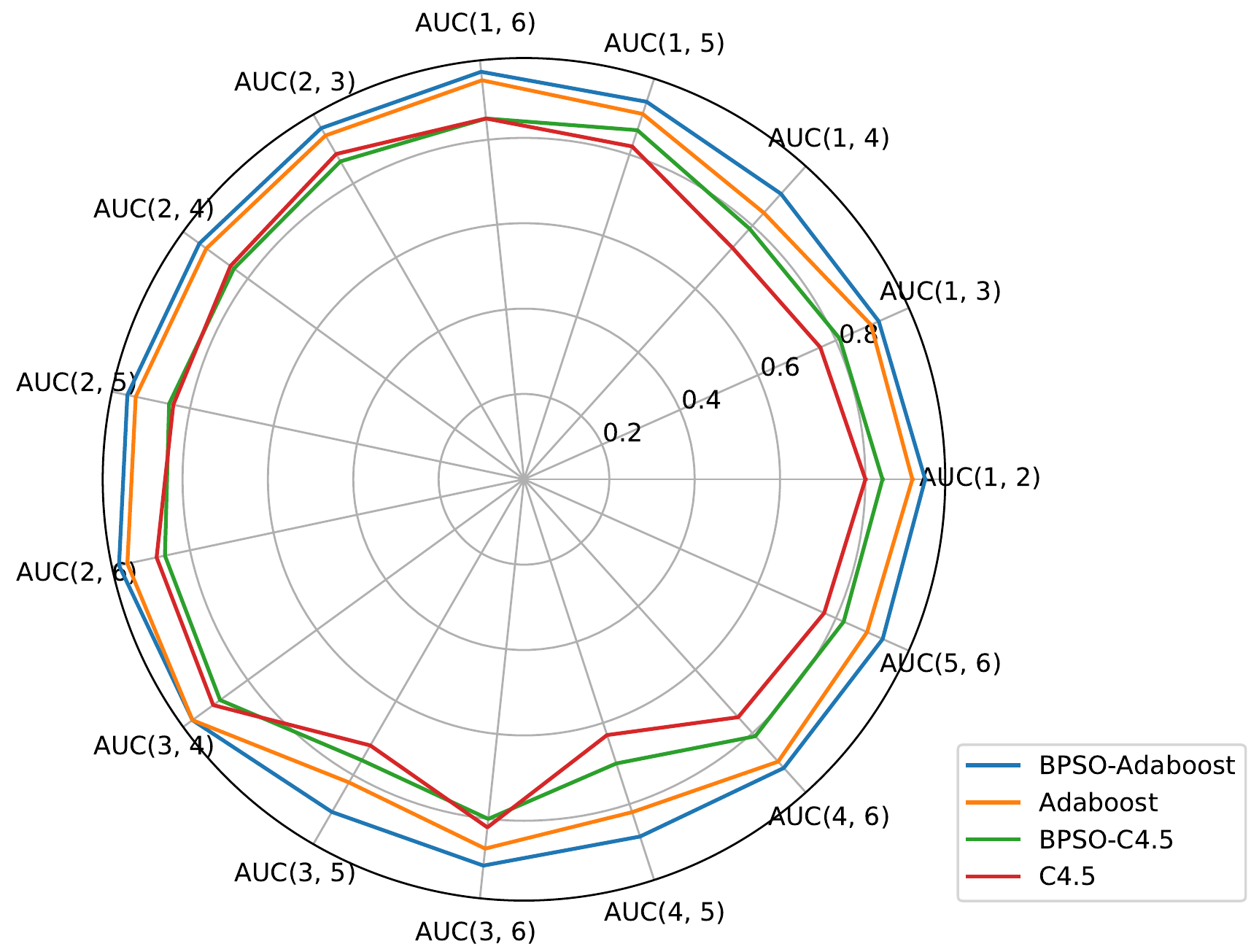}
  \caption{The AUC\_area of algorithms on full-feature model}
  \label{fig:full}
\end{figure}

\subsubsection{The Effect of Feature Selection and Ensemble Learning}
According to the experimental results, the relationship between two evaluation metrics remain consistent: higher AUC\_area values have higher accuracies. Overall, the performance of the BPSO-Adaboost algorithm is superior to the other three algorithms. The Adaboost.M1 algorithm performs much better than C4.5 over all categories, demonstrating that ensemble learning improves the performance of individual weak classifiers on the imbalanced data set. Besides, we found that algorithms with the BPSO algorithm exhibit better performance those without in terms of both overall accuracy and the accuracy of minority categories. It confirms our previous assumption that the BPSO algorithm can exclude redundant features, which reduces the noise in minority samples and improves the overall performance of the model. In conclusion, both feature selection and ensemble learning have positive contributions to our model; they play essential roles in classifying samples of minority categories, which are of particular concern in imbalanced data. 

\subsubsection{Robustness Comparison Between Our Approach and the State-of-the-art NLP-based Approach}
\begin{figure}[tb]
  \centering
  \includegraphics[width=0.87\linewidth]{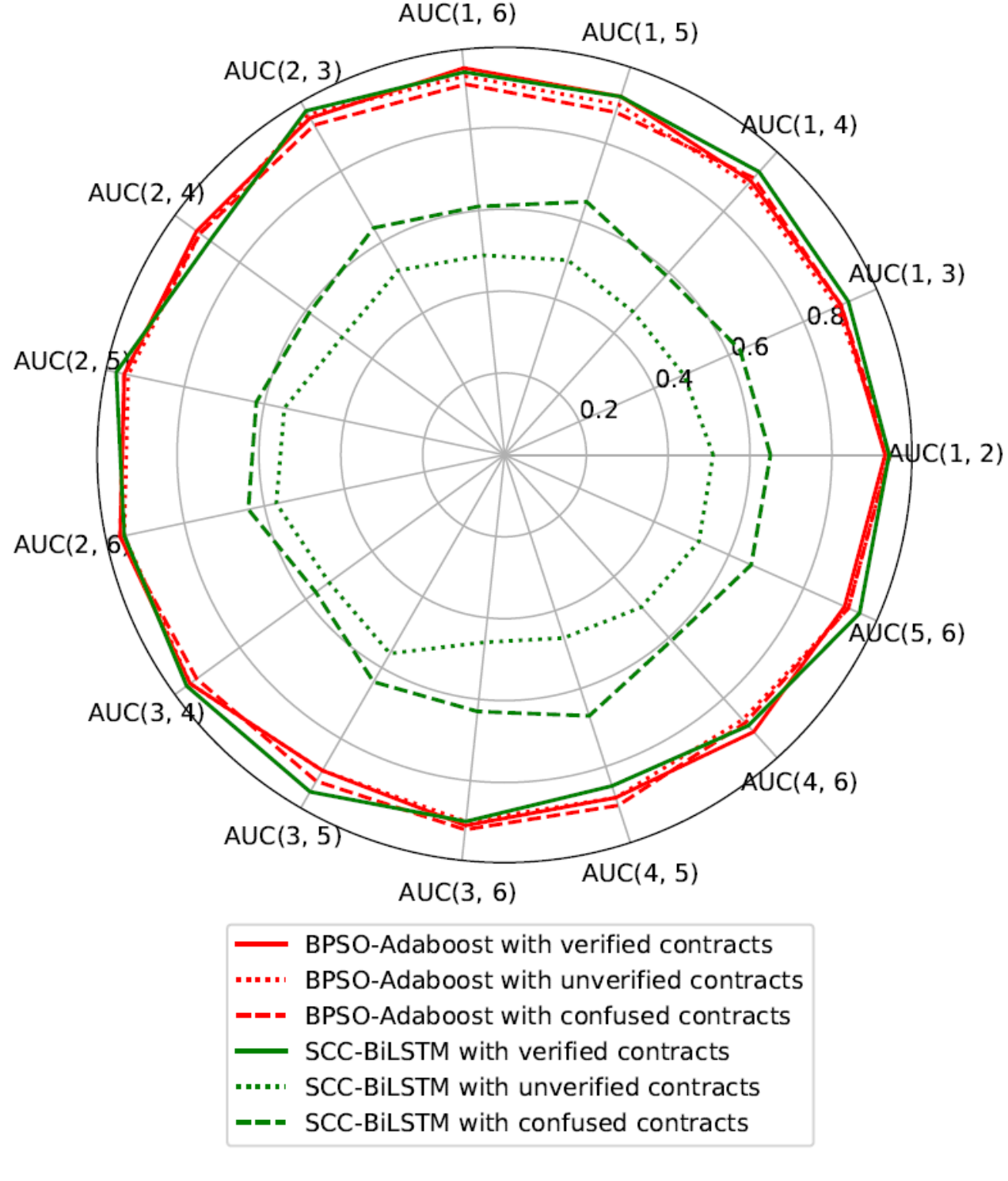}
  \caption{The AUC\_area of BPSO-Adaboost and SCC-BiLSTM under different test sets.}
  \label{fig:robust}
\end{figure}

Table~\ref{tab6} and Fig.~\ref{fig:robust} show the performance of our approach and a representative state-of-the-art NLP-based approach, the SCC-BiLSTM algorithm~\cite{tian2020smart}, under the different test sets mentioned in Section 5.2. For verified contracts, the SCC-BiLSTM algorithm has a slightly better performance than the proposed BPSO-Adaboost algorithm, but the results are very close. For unverified contracts, the BPSO-Adaboost algorithm retains its high performance; in contrast, the SCC-BiLSTM algorithm degenerates into a random classifier due to a lack of code features. The performance of the SCC-BiLSTM algorithm is also poor on adversarial examples, but the BPSO-Adaboost algorithm is barely affected. The reason is that the key features of the NLP-based approach are mainly semantic features that are not robust at all once attacked. For our bytecode-based approach, the attack on the source code also causes bytecode changes. However, the changes to equivalent opcodes happen only after \textbf{KECCAK256}~\cite{buterin2014next} because the opcodes of all the variable names, function names, and comments need to be computed as keccak-256 hashes. Thus the weights of opcodes after \textbf{KECCAK256} are significantly lower than opcodes that relate to core operations and have little impact on classification. These results indicate that our bytecode-based approach has similar performance with state-of-the-art NLP-based approach, and has much better robustness.

\section{Conclusion}
This paper proposes a novel bytecode-based classification approach designed to effectively classify smart contracts of blockchain platforms. Considering traditional classifiers have poor performance on imbalanced data sets, we use a feature selection method to reduce the noise of the samples as well as an ensemble learning approaach to improve the overall performance of the classifier. Comparative experiments prove the superiority of each element in our algorithm. The result of feature selection also reveals why the full-feature model has little improvement over the 0-day model. Compared with a state-of-the-art NLP-based approach, our bytecode-based approach provides good performance and offers two key advantages. First, it dramatically expands the application scenarios for which the classifier can be used (i.e., bytecode for open-source, non-open-source contracts). Second, our method can defend against semantic attacks. These results demonstrate our bytecode-based approach has better robustness than approaches that depend on contract source code. 

This paper focuses on demonstrating the bytecode-based model's advantages in the classification of smart contracts. In the future, we plan to further this study this problem from three aspects. The first is to extend the data set as the number of smart contracts proliferates. We will continually improve the model with more ground truth data, including promote the classification accuracy and increase the number of categories that the model can classify. However, some smart contracts belong to categories which are hard to classify clearly, such as contracts used for identification~\cite{8951253} or monitoring~\cite{8873576}. We expect there would be more detailed definitions of existing smart contracts. The second is to expand our smart contract classification model to other blockchain platforms. The potential targets are platforms with similar virtual machine architecture to EVM, e.g., Hyperleger Fabric~\cite{androulaki2018hyperledger}. Finally, we plan to explore more derivative functions based on the results of smart contract classification. For example, taking the popularity and gas efficiency of contracts into account to do top-k searching, or recommending specific categories of contracts to users based on their preferences.

\bibliographystyle{IEEEtran}
\bibliography{mybibfile.bib}

\end{document}